\documentclass[
reprint,
superscriptaddress,
amsmath,
amssymb,
aps,
prb,
floatfix
]{revtex4-2}
\usepackage{graphicx}
\usepackage{hyperref}
\usepackage{xcolor}
\usepackage{lipsum}

\begin{document}

\title{Learned Committors as Reaction Coordinates for Nucleation Rates}

\author{Hubert J. Naguszewski}
\thanks{Corresponding author}
\email{hubert.naguszewski@warwick.ac.uk}
\affiliation{Department of Physics, University of Warwick, Coventry, CV4 7AL, United Kingdom}
\author{David Quigley}
\affiliation{Department of Physics, University of Warwick, Coventry, CV4 7AL, United Kingdom}

\begin{abstract}
A central challenge in the analysis of first-order phase transitions is the identification of optimal reaction coordinates. In principle, the committor is the ideal choice; however, its computational cost has historically made it intractable. Here, we train a convolutional neural network ($p_B$-NN) as a proxy for the committor on brute-force committor labels and use it directly as the coordinate of a Markov state model. Applied to magnetisation reversal in the two-dimensional Ising model, $p_B$-NN reproduces brute-force nucleation rates across a range of thermodynamic conditions. The largest geometric cluster size also recovers accurate rates despite providing a poor pointwise predictor of the committor. These results demonstrate that an effective reaction coordinate for nucleation rate calculation must reliably separate the metastable and stable basins, but need not preserve the committor pointwise for every microstate. We stress that this distinction has direct implications for the choice of collective variable in rare-event simulations of nucleation more broadly.
\end{abstract}

\date{\today}

\maketitle

\section{Introduction}
\label{sec:introduction}

The dynamics of first-order phase transitions evolve via nucleation and growth, often on timescales inaccessible to direct computer simulation. Nucleation-rate calculations hence make use of advanced simulation techniques~\cite{blow2021sins} to capture the rare event corresponding to formation of a supercritical nucleus of the stable phase within a metastable parent phase, which then grows to macroscopic size.

A central challenge is the identification of low-dimensional descriptors, i.e. collective variables (CVs) or reaction coordinates of the process. In the spirit of classical nucleation theory (CNT), the natural choice is the size of clusters (or ``droplets") of the stable phase. It has long been recognised~\cite{langer_theory_1967} that reducing nucleation to a single degree of freedom is a simplification which may lead to loss of information. Despite this, nucleation rates based on this approximation can be remarkably accurate in some systems.

An archetypal example is magnetisation reversal in the two-dimensional nearest neighbour Ising model. A simple CNT-like CV is the number of spins in the largest connected cluster aligned with the external field, where spins are connected if they are mutually aligned. However, it is known that an alternative definition, specifically Fortuin--Kasteleyn clusters (FK)~\cite{AConiglio_1980, JANKE2004385}, is more consistent with coexistence thermodynamics~\cite{maibaum_phase_2008}, especially close to the critical temperature $T_c$~\cite{schmitz_monte_2013}. Cluster fluctuations and dynamics are also better characterised by a combination of size and perimeter rather than size alone~\cite{binder_clusters_1976, pan_dynamics_2004}.

In modern transition state theory~\cite{EricTPT}, it is recognised that the ideal CV is the committor $p_B$~\cite{krivov2013optimality,Peters_reaction_coordinates}. In the context of nucleation, this is the probability that a microstate will evolve to the thermodynamically stable phase B before returning to the parent phase A. The perfect CV will label only microstates with $p_B\approx 0.5$ as being part of the transition state ensemble. This leads to the ``histogram test"~\cite{Peters_histogram_test} in which the distribution of actual committor values is constructed for the transition state ensemble identified by the CV. A histogram tightly centred on $p_B=0.5$ represents a good choice of CV. In the context of the two-dimensional Ising model, it has been shown that a much tighter histogram is achieved by using a CV constructed from both cluster size \emph{and} perimeter~\cite{peters_obtaining_2006, moroni_interplay_2005}. In general, computing committors is expensive and usually restricted to this test. However, learning the committor has recently become tractable via the use of neural networks~\cite{bonati_data-driven_2020, fu_collective_2024, huang_predicting_2021, jung_aimmd_2023}.

In principle, a projection onto $p_B$ has dynamics optimally close to Markovian~\cite{peters_obtaining_2006}. By contrast, the dynamics of geometric cluster size can exhibit memory effects, particularly at short timescales or under spin-exchange dynamics\cite{kuipers_limitations_2010}. The latter has also been demonstrated for other lattice models in the context of free energy reconstruction based on a Fokker-Planck equation~\cite{Lifanov_Seeding}.

Despite these limitations, recent numerical studies using spin-flip dynamics have shown that CNT based on geometric cluster size can quantitatively agree with numerical results provided corrections for shape fluctuations are included, even in the presence of impurities~\cite{ryu_numerical_2010, mandal_nucleation_2021}. The numerical method used was forward flux sampling (FFS)\cite{AllenFFS}, which is relatively insensitive to the choice of CV and does not require Markovian dynamics. Their results are also consistent with earlier brute-force simulations~\cite{brendel_nucleation_2005}. This suggests that, while imperfect, geometric cluster size is often sufficient for accurate rate calculations in this model system. However, it remains unclear whether a more faithful CV would yield improved rate estimates, or whether the dominant nucleation pathways are already well captured by simple cluster size descriptors.

In this Letter, we push optimisation of the CV for magnetisation reversal in the two-dimensional Ising model to its logical limit by directly using the committor as the CV. Unlike approaches based on variational optimisation~\cite{kang_computing_2024, valsson_variational_2014, mitchell_committor_rates_2024}, we exploit massively parallel simulations on a modern GPU~\cite{Weigel_2017} to generate highly accurate ``ground truth" committors for $\sim 10^{3}$ microstates distributed along nucleating trajectories, and train a convolutional neural network as a proxy for $p_B$. We then construct Markov state models (MSMs) for the nucleation process, based both on our $p_B$ proxy ($p_B$-NN) \emph{and} the largest geometric cluster size (LGCS) for comparison. We pay particular attention to temperature and field values where the above literature suggests the geometric cluster definition leads to thermodynamically inconsistent sizes, and its dynamics are non-Markovian on short time scales.

\begin{figure*}[t]
    \centering
    \includegraphics[width=\linewidth]{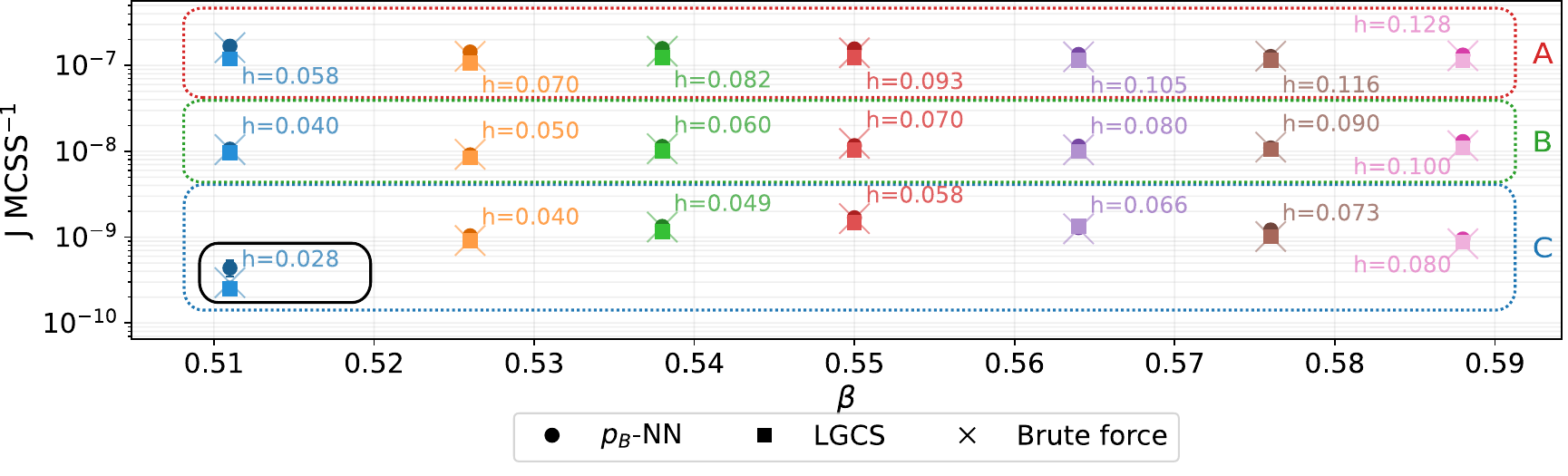}
    \caption{Nucleation rates across a range of $(\beta,h)$. Circles and squares denote rates from the learned committor coordinate ($p_B$-NN) and the largest geometric cluster size (LGCS) Markov State Models respectively. Brute-force magnetisation-reversal rates are denoted with crosses. Sections A, B, and C correspond to the fast, intermediate, and slow nucleation regimes. Error bars are smaller than the symbol sizes. The points outlined by the black box are excluded from further processing due to cluster percolation.}
    \label{fig:nucleation}
\end{figure*}

\section{Methods}
\label{sec:methods}

We begin by generating committor training data, i.e. many microstates of the Ising model, sampled along a brute-force nucleating trajectory, labelled with their associated committor value $p_B$. Since the nucleation rate of the Ising model is a function of field $h$ and the inverse temperature $\beta$, we can explore a range of nucleation rates. We simulate two-dimensional Ising lattices consisting of $L^2$ spin sites, where $L=64$, using single-spin-flip dynamics at the $(\beta,h)$ values shown in the Figure~\ref{fig:nucleation}. The points outlined by the black box are excluded from our study due to cluster percolation, details of which can be found within Section S8 of the Supplemental Material. We apply a positive magnetic field to configurations in the metastable parent phase with magnetisation $m\approx-1$ and hence the stable phase has magnetisation $m\approx 1$.
The metastable parent phase and stable phase boundaries used to determine whether a trajectory has reached either phase are subsequently defined. The peak in the LGCS probability distribution within the parent metastable phase defines the parent boundary, while the stable basin boundary is defined by $\text{LGCS}=0.5L^2$, beyond which the committor $p_B=1$ for any choice in the $(\beta,h)$ range which we study. For all of our $(\beta, h)$, the critical nucleus size is small enough such that it never interacts with itself across periodic boundaries.

We obtain brute-force estimates of the committor by launching $n=4096$ independent simulations from each of these microstates and recording the fraction of trajectories that reach the stable basin before returning to the parent basin. The choice of $n=4096$ satisfies our target ground truth committor label accuracy of $\pm 0.01$, and typically takes 1-10 seconds to evaluate $p_B$ for a single microstate on a modern GPU. For additional details of how microstates are selected and how errors are calculated we direct the reader to Section S2 of the Supplemental Material. These labelled microstates are then used to train a convolutional neural network (CNN)~\cite{lecun1998gradient, krizhevsky2012imagenet}, which serves as a proxy for $p_B$ and can be evaluated in milliseconds. The CNN takes as input a square grid of $L^2$ spins and passes them through three residual CNN blocks followed by two residual linear blocks with the single output value being passed through a final sigmoid activation layer. A closely related problem was studied by Huang et al.~\cite{huang_predicting_2021}, who trained a CNN to predict the probability that a microstate would nucleate in a fixed time window. Our significantly more accurate approach differs in that we learn the committor itself, translational symmetry is built into our network, and our model is larger (order $10^5$ against $10^4$ parameters). For more details on the neural network architecture, we direct the reader to Section S6 of the Supplemental Material. Each thermodynamic condition is treated with a separately trained model using an average of 2000 labelled microstates, which is sufficient to result in a $p_B$ proxy with an average root-mean-square error (RMSE) on its prediction of $\pm 0.01$, i.e. limited only by the accuracy of the training data.

Once we have obtained $p_B$-NN at a range of temperatures and fields, we create separate MSMs using $p_B$-NN and the LGCS as the Markov state CVs for each of our $(\beta,h)$. The MSM transition matrix is estimated at lag time $\tau$ by initiating trajectories of length $\tau$ from microstates distributed along the chosen coordinate, and rates are obtained from the mean first-passage time from the parent basin to the stable basin. The two reaction coordinates are discretised independently. For the LGCS MSMs we use 20 states between the basins, which we find sufficient to obtain converged rates from the implied-timescale criterion---explained later within this Letter. For the $p_B$-NN proxy we use 40 states since this coordinate is itself a committor estimate, and using 40 states gives a resolution of $\sim$2.5\% in the committor per state---fine enough to discriminate neighbouring states. In order to ensure that the MSM rates are converged with respect to lag time, we select the shortest lag time at which the longest implied timescale of the MSM has plateaued, and report the corresponding rate. The timescale is defined as the largest eigenvalue not equal to 1 of the MSM transition matrix. In order to compare the rates obtained from the $p_B$-NN and the LGCS MSMs, we compare them to rates obtained via brute-force simulation which---where applicable---are in agreement with the values obtained by Brendel et al.~\cite{brendel_nucleation_2005}. The details of the brute-force calculations can be found in Section S2 Supplemental Material.

For further comparison, we also use the mean largest FK cluster size as a pointwise committor predictor. This is obtained by constructing an FK cluster using the connecting probability $p=1 - e^{-2\beta}$, which dictates whether neighbouring mutually aligned spins form a connected cluster. We perform 32 such FK cluster calculations per Ising microstate to determine a mean largest FK cluster size.

Finally, to allow for analysis of the internals of $p_B$-NN, we employ saliency maps~\cite{Simonyan2014saliency}. These maps are constructed by computing the gradient of the predicted committor with respect to each site in the input configuration, thereby identifying the spins that most strongly influence the $p_B$-NN prediction.

\section{Results and Discussion}
\label{sec:results_discussion}

We first benchmark MSM nucleation rates constructed along $p_B$-NN and LGCS against direct brute-force rate calculations, before examining the pointwise committor accuracy of $p_B$-NN, and of fitted sigmoid functions of LGCS and FK to yield CVs which span the range [0,1]. Comparing nucleation rates from $p_B$-NN and LGCS reveals a disconnect between rate and pointwise committor prediction accuracy, which we examine in detail. Finally, the saliency maps confirm that $p_B$-NN has learned physically meaningful features of the Ising microstates, and highlight the flaws associated with LGCS as a proxy for the committor.

To establish $p_B$-NN as a practical reaction coordinate, MSM rates constructed along $p_B$-NN and LGCS are compared with brute-force magnetisation-reversal rates across a range of $(\beta,h)$ in Figure~\ref{fig:nucleation}, which is divided into three nucleation regimes: fast (A), intermediate (B), and slow (C). Both $p_B$-NN and LGCS MSM rates are consistent with their brute-force counterparts across all regimes. Within the fast regime, however, LGCS appears to perform worse compared to both $p_B$-NN and the brute-force rate values, more details of which can be found in Section S7 of the Supplemental Material.

\begin{figure}[t]
    \centering
    \includegraphics[width=\linewidth]{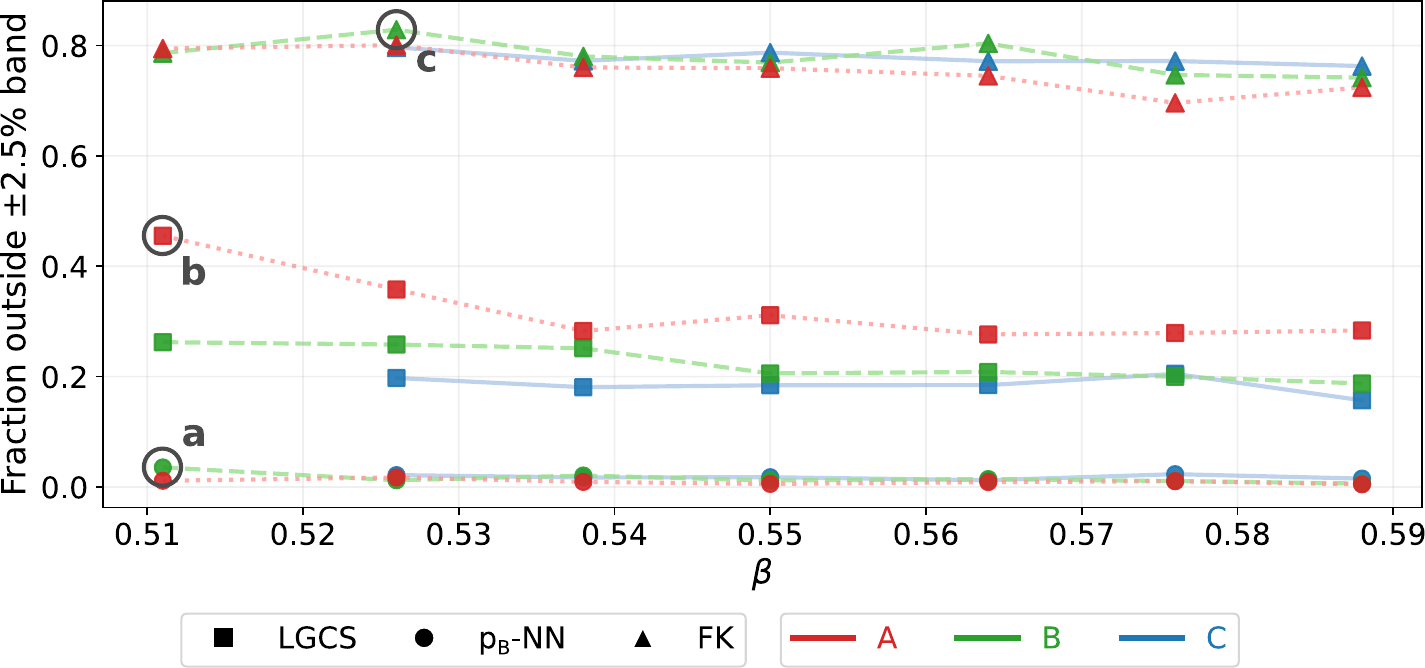}
    \caption{Fraction of committor predictions outside a $\pm 2.5\%$ tolerance band. Circles, squares, and triangles denote the learned committor coordinate ($p_B$-NN), the largest geometric cluster size (LGCS), and the Fortuin--Kasteleyn largest-cluster size (FK), respectively. Red, green, and blue points correspond to the fast, intermediate, and slow nucleation regimes of Figure~\ref{fig:nucleation}. Points labelled \textbf{a}, \textbf{b}, and \textbf{c} mark the thermodynamic conditions at which each coordinate exhibits its largest fraction outside the band; the corresponding committor predictions are shown in Figure~\ref{fig:committor-scatter}.}
    \label{fig:diff-fraction}
\end{figure}

Having established rate agreement, we now turn to the underlying pointwise committor predictions. Figure~\ref{fig:diff-fraction} shows the fraction of committor predictions lying outside a $\pm 2.5\%$ tolerance band for $p_B$-NN, LGCS, and FK, organised by the same three nucleation regimes used in Figure~\ref{fig:nucleation}. For $p_B$-NN, the fraction outside the tolerance band remains below 5\% across all conditions, demonstrating that the learned coordinate is an accurate pointwise predictor of the committor. By contrast, both LGCS and FK exhibit substantially larger fractions, indicating significant loss of microscopic committor information when the dynamics are projected onto a single scalar cluster-size coordinate.

\begin{figure}[t]
    \centering
    \includegraphics[width=\linewidth]{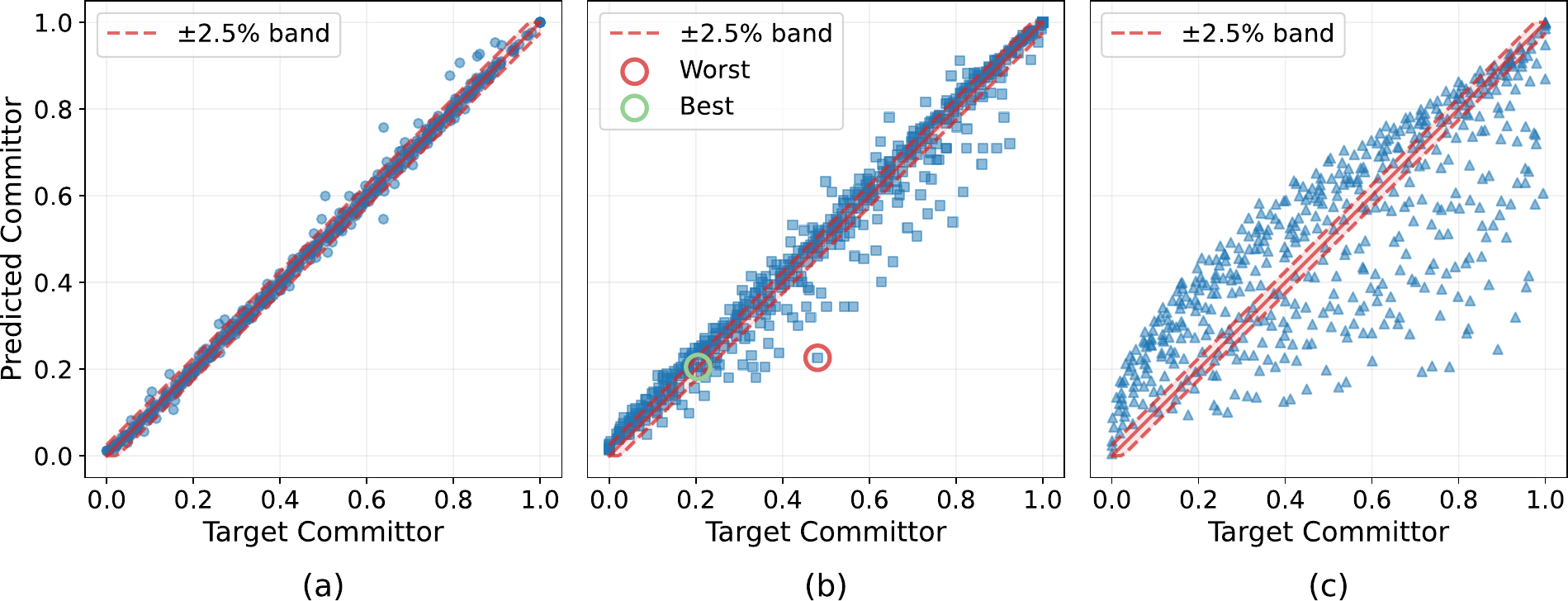}
    \caption{Predicted committor values compared with brute-force target values for $p_B$-NN (a), LGCS (b), and FK (c). The values of $(\beta,h)$ are $(0.511, 0.040)$, $(0.511, 0.058)$, and $(0.511, 0.028)$, respectively. The mappings from reaction coordinate to committor in panels (b) and (c) are obtained by fitting a sigmoid.}
    \label{fig:committor-scatter}
\end{figure}

These differences are shown in Figure~\ref{fig:committor-scatter}, where for each coordinate we select the $(\beta, h)$ with the largest fraction outside the tolerance band---points labelled (a), (b), and (c) in Figure~\ref{fig:diff-fraction}---and compare predicted committor values directly against the ground truth committor labels. Panels (a) and (b) of Figure~\ref{fig:committor-scatter}, corresponding to $p_B$-NN and LGCS, reveal a substantial difference in pointwise accuracy: $p_B$-NN predictions remain tightly distributed about the target committor, whereas LGCS predictions exhibit broader scatter. Despite this, both coordinates produce nucleation rates in agreement with brute force calculations (Figure~\ref{fig:nucleation}). This is the central observation of the present work: pointwise committor accuracy and rate accuracy are distinct requirements, and the latter can be satisfied even by a coordinate that fails the former.

Closer inspection of panel (b) shows why. Although individual LGCS predictions exhibit significant variance, the average prediction at fixed committor remains approximately correct. In other words, while geometric cluster size cannot uniquely determine the committor of an individual microstate, it does capture the dominant statistical trend of the nucleation process, and this is sufficient for an MSM constructed along it to recover accurate rates. We have also considered whether a more accurate prediction of pointwise committor values requires less averaging for the MSM calculation, however we were not able to resolve any such difference in our calculation.

The FK coordinate, seen in panel (c), performs substantially worse than either alternative as a pointwise predictor. Although FK clusters are more consistent with coexistence thermodynamics than geometric clusters~\cite{maibaum_phase_2008, schmitz_monte_2013}, this thermodynamic consistency does not translate into modelling the committor. We attribute this to the probabilistic dilution of FK clusters, which can fragment physically connected clusters, particularly in the highly non-spherical cluster microstates found in two-dimensional Ising model nucleation. We emphasise that FK clusters were constructed to be consistent with coexistence thermodynamics, not as nucleation CVs; their poor pointwise performance here is therefore consistent with, rather than contrary to, the literature.

\begin{figure}[t]
    \centering
    \includegraphics[width=0.85\linewidth]{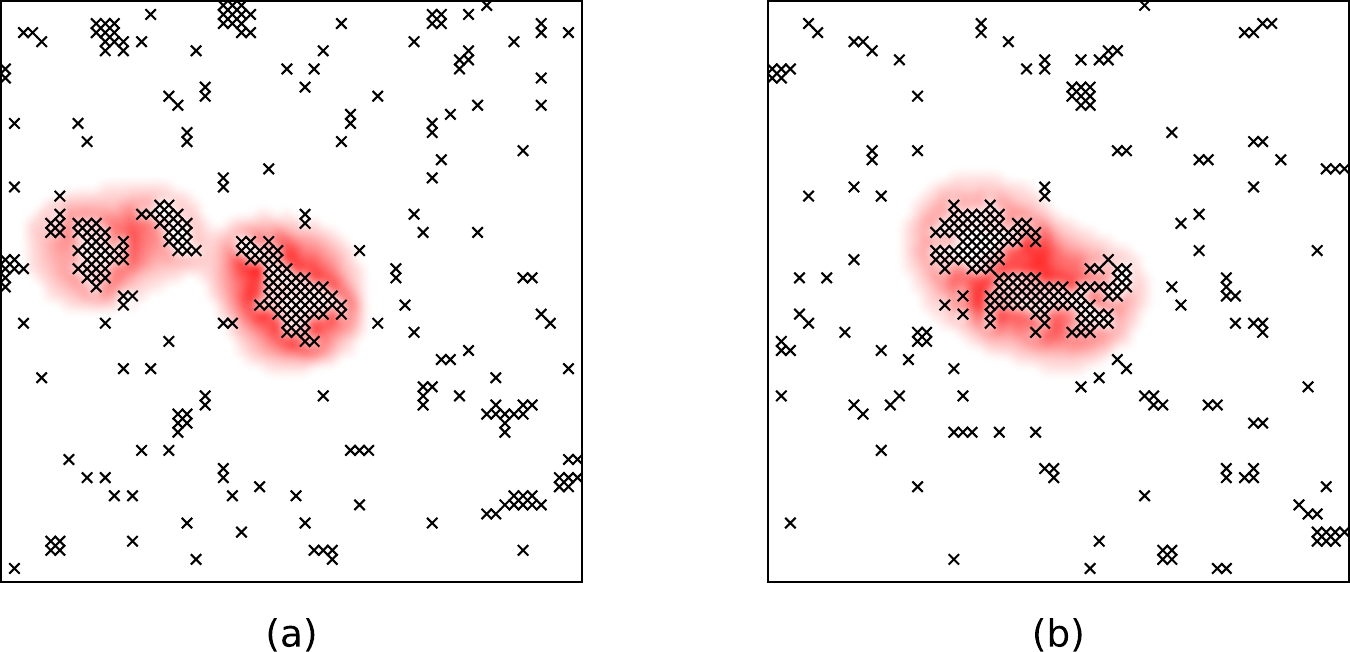}
    \caption{Saliency maps of $p_B$-NN for two representative configurations drawn from the LGCS-worst condition of Figure~\ref{fig:committor-scatter}(b). Red regions correspond to the saliency map with darker shades indicating a larger impact on the final predicted $p_B$ value. Panel (a) shows a configuration for which LGCS produces a near-correct committor estimate ($p_B^\text{true}=0.205\pm 0.006$, $p_B^\text{LGCS}=0.205 \pm 0.137$, $p_B^\text{NN}=0.204 \pm 0.017$). Panel (b) shows a configuration for which LGCS fails ($p_B^\text{true}=0.481\pm 0.008$, $p_B^\text{LGCS}=0.226 \pm 0.137$, $p_B^\text{NN}=0.487 \pm 0.019$).}
    \label{fig:saliency}
\end{figure}

To verify that $p_B$-NN has learned physically relevant features, we examine saliency maps computed from the input gradients of the trained network. Figure~\ref{fig:saliency} shows representative results for the two configurations drawn from the LGCS worst condition of panel (b) in Figure~\ref{fig:committor-scatter}: one for which LGCS happens to produce a correct committor estimate (panel (a) of Figure~\ref{fig:saliency}), and one for which LGCS vastly underestimates the committor (panel (b) of Figure~\ref{fig:saliency}). In both cases, the dominant contributions to the $p_B$-NN prediction are concentrated around clusters of up-spins and their boundaries, confirming that the network has learned physically meaningful features of the Ising configurations.

More importantly, the configuration in panel (b) of Figure~\ref{fig:saliency} is representative of the multi-cluster paths mentioned above. LGCS, which considers only the largest cluster, estimates a committor value far from the brute-force target, whereas $p_B$-NN---which considers all clusters, as seen in the highlighted saliency regions---recovers the target value to within statistical uncertainty. This difference in prediction comes about not from the mere presence of multiple clusters, but from cluster proximity which provides a pathway to the stable state via a merging of clusters.

Despite the clear pointwise advantage of $p_B$-NN, we stress that the LGCS MSM still produces rates close to brute-force values across all regimes. The pointwise failures, like the one seen in panel (b) of Figure~\ref{fig:saliency}, do not propagate into rate errors, which reinforces our central observation: an effective reaction coordinate for rate calculations must reliably separate the metastable and stable basins, but need not preserve the committor pointwise for every microstate.

\section{Summary and Conclusions}
\label{sec:conclusions}

The central finding in this Letter is that pointwise accuracy of a committor proxy and accuracy of the nucleation rates derived from it are distinct requirements. In the two-dimensional Ising model, MSMs constructed along the learned committor coordinate $p_B$-NN give nucleation rates consistent with brute-force magnetisation-reversal calculations, demonstrating that a neural network committor proxy is suitable not only as a diagnostic quantity but as the coordinate on which a kinetic model is built. At the same time, the largest geometric cluster size---which can fail substantially as a pointwise committor predictor---still yields rates in agreement with brute force.

Saliency analysis of the trained $p_B$-NN networks confirms that the dominant contributions to the predicted committor are concentrated around clusters of up-spins and their boundaries, indicating that the network has learned features of clear physical relevance. We deem such interpretability checks important, particularly when machine-learned coordinates are used as the basis for kinetic modelling, since they directly indicate which physical features are most relevant for committor estimation.

These findings have direct implications for transition state theory in nucleation contexts. They suggest that the well-known shortcomings of geometric cluster size as a pointwise committor proxy in off-lattice simulations ~\cite{peters_obtaining_2006, kuipers_limitations_2010} should not preclude its use for accurate rate estimation in this particular system. Conversely, the success of $p_B$-NN as an MSM coordinate opens a clear route to systematic improvement of nucleation rate calculations in systems where geometric cluster size is known to fail, such as those with strong shape fluctuations or competing crystalline polymorphs. Natural extensions include the construction of committor models which transfer across thermodynamic conditions, and application of the present strategy to off-lattice and multi-component nucleation problems where accurate ML potentials are now available~\cite{piaggi_ice_2022}.

\section*{Author Contributions}
H.J.N. and D.Q. conceived of the approach. H.J.N. performed the simulations, committor calculations, neural network training, MSM construction and analysis, and data analysis. D.Q. implemented the GPU-accelerated Ising model generator. Both authors interpreted the results, drafted and revised the manuscript, and approved the final version.

\section*{Data Availability}
The data supporting the findings of this study are freely available at the following DOI/URL: \href{https://doi.org/10.5281/zenodo.17629514}{https://doi.org/10.5281/zenodo.17629514} (will be made public upon journal acceptance).

\begin{acknowledgments}
H.J.N. acknowledges a funded studentship within the UK Engineering and Physical Sciences Research Council (EPSRC) Centre for Doctoral Training in Modelling of Heterogeneous Systems, Grant EP/S022848/1. Calculations were performed using the Sulis Tier 2 HPC platform hosted by the Scientific Computing Research Technology Platform (SCRTP) at the University of Warwick. Sulis is funded by EPSRC Grant EP/T022108/1 and the HPC Midlands+ consortium. Additional computing facilities were provided by the Scientific Computing Research Technology Platform at the University of Warwick.
\end{acknowledgments}

\end{document}